\documentclass[aps,prl,reprint,groupedaddress,twocolumn]{revtex4-1}

\usepackage{graphicx}
\usepackage{amsfonts}
\usepackage{mathrsfs}
\begin{document}

\title{Quantum Kinetic Theory of Collisionless Superfluid Internal Convection}

\author{Lukas Gilz}
\email[]{lgilz@rhrk.uni-kl.de}
\author{James R. Anglin}

%\homepage[]{Your web page}
%\thanks{}
%\altaffiliation{}
\affiliation{\mbox{State Research Center OPTIMAS and Fachbereich Physik,} \mbox{Technische Univerit\"at Kaiserslautern,} \mbox{D-67663 Kaiserslautern, Germany}}
\affiliation{Graduate School Materials Science in Mainz, Staudinger Weg 9, D-55128 Mainz, Germany}
\pacs{}
\date{\today}

\begin{abstract}

Superfluids can transport heat via simultaneous opposite flows of their spatially interpenetrating condensate and thermal components. While this \textit{internal convection} is usually described within Landau's phenomenological two fluid hydrodynamics, we apply quantum kinetic theory to a dilute Bose gas held beween thermal reservoirs at different temperatures, and show that the phenomenon also appears in collisionless kinetic regimes, and should be directly observable in currently feasible experiments on trapped ultracold vapors.
\end{abstract}

\maketitle

Because the dynamical simplicity of dilute ultracold gases allows us to distinguish essential mechanisms from the complex details that abound elsewhere, they are ideal many-body systems in which to probe the still poorly understood foundations of non-equilibrium statistical mechanics. Non-equilibrium time evolution in trapped quantum gases is a topic of high current interest \cite{thermalize, coherence, mott}. Besides investigating non-equilibrium time evolution, however, cold gas experiments can also impose a wide range of spatial inhomogeneities, offering the complementary opportunity to study quasi-stationary transport phenomena. Yet experimental samples are typically so small and dilute that, although the gas does equilibrate by scattering and exhibits nonlinear mean field dynamics, kinetics is `collisionless,' in the sense that the mean free path of quasiparticles is larger than the system size. In the consequent absence of local equilibration, most forms of transport cannot appear. But in this Letter we show that an unusual form of heat transport, known as \textit{internal convection}, should be observable even in a collisionless Bose-condensed gas. 

Internal convection has been observed in superfluid Helium \cite{tilley}, and is usually described within Landau's phenomenological two-fluid theory  \cite{landau}: under a temperature gradient, the thermal and Bose-condensed fractions flow through each other, in opposite directions, with no net flux of mass. As if to rebut the 19th century conclusion that cold is only lack of heat, condensate flows as a flux of cold, from cooler regions to hotter.  This striking yet simple form of transport is a convection in the sense that it involves motional separation of the fluid, but in Fourier rather than physical space. While identifying superfluid and normal components in Helium can be subtle, the relative motion of condensate and thermal fractions in a dilute Bose gas is determined straightforwardly, according to the standard Penrose-Onsager definition \cite{pen-ons}, from the single-particle reduced density matrix. Direct observation of these two fractions in time of flight expansion is a standard thermometry technique in experiments. 

Since internal convection does not involve spatial separation, we will for simplicity consider a quasi-one-dimensional sample. Although one-dimensional many-body physics can differ dramatically from that in higher dimensions, for finite samples there exists a (quasi-\nolinebreak) condensate regime \cite{Petrov, liebliniger} in which exactly similar mean field and quasiparticle theory governs both cases. This has been confirmed experimentally, for example in tests of Gross-Pitaevskii soliton dynamics in one-dimensional systems \cite{soliton}. Our results are therefore not special to one dimension, but generic. Moreover we derive them in the collisionless limit, but hydrodynamic internal convection is already well known. And we estimate readily observable flow speeds, approaching the range of mm/s under currently attainable conditions. Our conclusion is that internal convection is a robust phenomenon that should be observable even well beyond the realistic regime that we consider.

\begin{figure}
\includegraphics[scale=0.22]{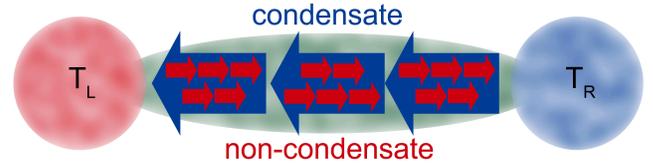}
\label{fig:sys}
\caption{(color online) Internal convection: A condensed Bose gas in contact with heat baths of different temperatures $T_L > T_R$.  Condensate flows towards the hotter reservoir, non-condensate in the opposite direction.}
\end{figure}

Heat transport in a dilute Bose gas under a temperature gradient has indeed already been analyzed quantitatively in \cite{zaremba1999}, though without explicitly identifying internal convection. Since collisionless gas does not support local equilibrium with a smoothly varying temperature, however, we take a different approach. Rather than assuming a temperature gradient, we let two reservoirs of thermal gas, with unequal temperatures $T_{L}$, $T_R$, interact with a Bose-condensed system cloud in different regions, as sketched in Fig. 1.  Such a configuration could be achieved by trapping atoms of two different elements, or in a spinor condensate with spin-flipping collisions suppressed \cite{suppress}.  One species of atoms serves as the system, and the other populates the reservoirs, with the three samples spatially localized by species-dependent external potentials \cite{species}. Species conservation prevents particle exchange between system and reservoirs, but collisions will still permit energy exchange \cite{anglin1999}.  Solving for the system steady state reveals non-equilibrium transport that includes internal convection.

We identify this steady state by finding the time-independent solution to a quantum kinetic master equation for the full density operator $\hat{\rho}$ of the system gas. We derive this master equation by tracing over the reservoir sectors of the total Hilbert space, using the standard Born-Markov elimination method of quantum optics as applied to quantum gas kinetics by Gardiner and Zoller {\it et al.} \cite{qkt1, qkt2, qkt3, qkt4, qkt5}. For technical convenience we consider a discretized model with periodic boundary conditions: a Bose-Hubbard ring of $2M$ sites, with the two reservoirs coupled only to the diamatrically opposite sites $m=0,M$, applying the reservoir-plus-localized-mode theory of \cite{anglin1997,ruostekoski1998}. We then take a continuum limit of our model to describe continuous condensates. 

The Born-Markov master equation presumes that the local atomic destruction operator $\hat{a}_{m}$ at each Bose-Hubbard site $m$ has the interaction picture evolution
\begin{eqnarray}
\hat{a}_m&=&\sum_{\omega}\hat{\alpha}_m(\omega)e^{-i\omega t} \label{localx}
\end{eqnarray}
for some set of eigenfrequencies $\omega$ and operator-valued Fourier co-efficients $\hat{\alpha}_m(\omega)$. Assuming that the reservoir coherence time is long enough to resolve all the system frequencies $\omega$, we obtain \cite{tbppra}
\begin{eqnarray}\label{master1}
\frac{d}{dt}\hat{\rho}	&=&\frac{i}{\hbar}\!\left[\hat{\rho},\hat{H}\right]-\gamma\!\!\!\sum_{m=0,M\atop \omega, \omega'}\hat{\mathcal{M}}_m(\omega,\omega')\hat{\rho}\\	\label{eq:meg}
\hat{\mathcal{M}}_{m}(\omega,\omega')\hat{\rho}&=& R\bigl({\beta_m}(\omega-\omega')\bigr)[\hat{\rho}\,\hat{\alpha}_m^\dagger(\omega')\hat{\alpha}_m(\omega),\hat{a}^\dagger_m\hat{a}_m]\nonumber\\
&&- R\bigl({\beta_m}(\omega'-\omega)\bigr)[\hat{\alpha}_m^\dagger(\omega')\hat{\alpha}_m(\omega)\hat{\rho},\hat{a}^\dagger_m\hat{a}_m] \nonumber\\
R(z) &=& \left\{\begin{array}{ll}1, & z < 0\\e^{-z}, & z \geq 0\end{array}\right.\nonumber
\end{eqnarray}
where $\hat{H}$ is the system Hamiltonian, $\gamma$ is the effective scattering rate of the reservoirs (taken as the same for both of them, for simplicity), and $\beta_{0,M} = 1/(k_{B}T_{L,R})$ the inverse temperature of the respective reservoir. The non-Hamiltonian terms in our master equation (\ref{master1}) provide energy gain and loss, weighted by appropriate thermal factors in $R(z)$, and with site-dependent phases contained in the Fourier co-efficients $\hat{\alpha}_{0,M}$. 

For $\hat{H}$ we take a one-dimensional Bose-Hubbard model, which can represent bosons trapped in a lattice potential or, in the continuum limit, a bulk sample. This discretization will also facilitate comparison with transport problems in spin chains \cite{spinchains2} and strongly interacting quantum gases \cite{spinchains1}.
\begin{eqnarray}
\frac{\hat{H}}{\hbar}=\sum_m\left[\frac{\kappa \hat{a}_m^{\dagger 2}\hat{a}_m^2}{4\bar n}-\mu\hat{a}_m^{\dagger}\hat{a}_m-J(\hat{a}^{\dagger}_m\hat{a}_{m+1}+\mathrm{H.c.})\right]\label{BH}\end{eqnarray}  
Here $J$ is the tunneling rate between neighboring lattice sites, $\bar{n}=N/(2M)$ is the average site occupation, and $\kappa/(4\bar{n})$ is the on-site interaction energy. The system's chemical potential $\mu$ may be tuned for convenience, since $\hat{N}=\sum_{m}\hat{a}_{m}^\dagger\hat{a}_{m}$ commutes with $\hat{H}$. We will consider an eigenstate of $\hat{N}$, with eigenvalue $N$.

The master equation (\ref{master1}) with Bose-Hubbard Hamiltonian (\ref{BH}) admits no general analytic solution. But in the strongly (quasi-)condensed case where almost all of the $N\gg2M$ particles occupy a single mode, so that $\bar{n}\gg1$ while $\kappa \lesssim J$, we can solve for the steady state $\hat{\rho}$ analytically within a number-conserving Bogoliubov expansion. We begin with the Holstein-Primakoff transformation
\begin{eqnarray}
\hat{a}_m	=\frac{\hat{A}}{\sqrt{2M}}  \left(\sqrt{\frac{\hat{N}\!-\!\sum_{k}\!\hat{b}^{\dagger}_k\hat{b}_k}{\hat{N}}}+	\frac{1}{\sqrt{\hat{N}}} \!\sum_{k=1}^{2M-1}e^{i\frac{\pi k m}{M}}\hat{b}_k\right)\label{eq:bogoliubov}\;\;
\end{eqnarray}
which leaves $\hat{A}$ as a canonical lowering operator for $\hat{N} \equiv \hat{A}^{\dagger}\hat{A}$, while the $\hat{b}_{k}$ are canonical bosonic destruction operators that all commute with $\hat{A}$ (and so with $\hat{N}$). We then assume a total number eigenstate $\hat{N}\to N$, and expand $\hat{H}$ in inverse powers of average site occupation $\bar{n}$. With a subsequent Bogoliubov transformation,
\begin{eqnarray}\label{boguv}
	\hat{b}_k=u_{k}\hat{c}_k+v_{k}\hat{c}_{2M-k}^{\dagger}
\end{eqnarray}
for canonical quasi-particle operators $\hat{c}_k$, and coefficients $u_{k}\pm v_{k}=\left(\omega_{k}/\Omega_{k}\right)^{\pm1/2}$, $\hat{H}$ reads:
\begin{eqnarray}
\hat{H}&=&\sum_{j=0}\hat{H}_{j}\,{\bar{n}^{-j/2}}, \\
\hat{H}_{0}&=&\sum_{k=1}^{2M-1}\hbar\Omega_k\hat{c}_k^{\dagger}\hat{c}_k\label{eq:BHbog}
\end{eqnarray}
%\hat{H}_{1}&=&2\kappa\sum_{k,l}\left(\hat{b}_k^{\dagger}\hat{b}_l\hat{b}_{k-l}+H.c.	\right)\label{eq:addham}\\
Herein, the eigen-frequencies $\Omega_{k}=\sqrt{\omega_{k}(\omega_{k}+\kappa)}$ contain the eigen-frequencies $\omega_{k}:=4J\sin^{2}\frac{\pi k}{2M}$ of the non-interacting model. 

With this Bogoliubov expansion we solve for $\dot{\hat\rho}=0$ by perturbing in $1/\bar{n}$ and $\gamma/\Omega_{1}$.  Assuming that although the reservoir coherence time is long, it is not long enough to resolve the very small frequency corrections from $\hat{H}_{1}$, we base our Born-Markov interaction picture on $\hat{H}_{0}$, so that we have $\omega\to\pm\Omega_{k}$ (or 0) in (\ref{localx}) and (\ref{master1}), with
\begin{eqnarray}\label{QPME}
	\hat{\alpha}_{m}(\Omega_{k}) &=& u_{k}e^{i\frac{\pi km}{M}}\frac{\hat{A}}{\sqrt{2MN}}\hat{c}_{k}\nonumber\\
\hat{\alpha}_{m}(-\Omega_{k}) &=& v_{k}e^{i\frac{\pi km}{M}}\frac{\hat{A}}{\sqrt{2MN}}\hat{c}^{\dagger}_{2M-k}\\
\hat{\alpha}_{m}(0) &=& \frac{\hat{A}}{\sqrt{2M}}\sqrt{1-\frac{\sum \hat{b}_{k}^{\dagger}\hat{b}_{k}}{N}}\nonumber
\;.
\end{eqnarray}
Inserting this with (\ref{eq:bogoliubov}) and (\ref{boguv}) into (\ref{master1}), we find that our master equation describes the particle-conserving excitation and damping of Bogoliubov quasi-particle modes in the Bose-Hubbard system, due to scattering of reservoir particles from sites $m=0$ and $M$. 

Expanding $\hat{H}$, $\hat{\mathcal{M}}_{m}$ and $\hat{\rho}$ in $1/\sqrt{\bar{n}}$ and in $\gamma$,
 \begin{eqnarray}
\hat{\rho}&=&\sum_{i,j=0}\hat{\rho}_{ij}(\gamma\bar{n})^{i}\,{\bar{n}^{-j/2}}\;,\label{eq:rho}\\
\hat\mathcal{M}_m(\omega,\omega')&=&\gamma\bar{n}\sum_{j=0}\hat{\mathcal{M}}_{mj}\,\bar{n}^{-j/2}\;,
\end{eqnarray}

leads to the following set of equations:
\begin{eqnarray}
\left[\hat{\rho}_{00},\hat{H}_{0}\right]\!\!&=&\!0\label{eq:order0}\\
\left[\hat{\rho}_{01},\hat{H}_{0}\right]\!\!&=&\!-\!\left[	\hat{\rho}_{00},\hat{H}_1	\right]\label{eq:order1}\\
\left[\hat{\rho}_{10},\hat{H}_{0}\right]\!\!&=&\!i\hbar\!\!\!\!\sum_{m=0,M}\!\!\!\!\hat{\mathcal{M}}_{m0}\hat{\rho}_{00}\label{eq:order2}\\
\left[\hat{\rho}_{11},\hat{H}_{0}\right]\!\!&=&\!i\hbar\!\!\!\! \sum_{m=0,M}\!\!\!\!\left(\hat{\mathcal{M}}_{m0}\hat{\rho}_{01}+\hat{\mathcal{M}}_{m1}\hat{\rho}_{00}\right)\!-\!\left[	\hat{\rho}_{10},\hat{H}_1	\right]\label{eq:order3}\;\;\;\;
\end{eqnarray}
Equations (\ref{eq:order0})--(\ref{eq:order3}) include no terms of order $\bar{n}^2$ or $\bar{n}^{3/2}$. Such terms are in principle possible, but vanish identically, showing that although internal convection is a coherent effect, it is not superradiant.

As usual in a perturbative master equation, the zeroth order equation (\ref{eq:order0}) determines $\hat{\rho}_{00}$ to be diagonal in the $\hat{H}_{0}$ basis, but does not fix its diagonal entries. Rather, the diagonal elements of the left side of (\ref{eq:order2}) vanish identically, determining $\hat{\rho}_{00}$ through the diagonal terms of $\hat{M}_{m0}$. This yields a canonical distribution for each mode, but with a temperature that depends on $\Omega_{k}$ (albeit weakly for small temperature difference):
\begin{eqnarray}\label{rho00}
\hat{\rho}_{00}&=&\frac{1}{Z}\prod_{k}\left[\frac{e^{-\beta_L\hbar\Omega_k}+e^{-\beta_R\hbar\Omega_k}}{2}\right]^{\hat{n}_k}
%&=&\frac{1}{Z}\prod_{k}e^{-\beta\hbar\Omega_k\hat{n}_k}\cosh\left(\Delta\beta\hbar\Omega_k\right)^{\hat{n}_k}
\end{eqnarray}

The solution to the next higher order equations can be expressed in the form $\hat{\rho}_{ij}=\hat{\rho}_{00}\hat{C}_{ij}$. The detailed evaluation of the operator-valued factors $\hat{C}_{ij}$ is lengthy but straightforward and therefore shown elsewhere \cite{tbppra}. Here it suffices to note that $\hat{C}_{01}$ and $\hat{C}_{11}$ include only combinations of odd numbers of quasi-particle creation and annihilation operators $\hat{c}_k^{\dagger}$ and $\hat{c}_k$,  whereas $\hat{C}_{10}$ consists solely of even-numbered combinations. 
%\begin{eqnarray}
%\hat{\rho}_{01}&=
%&\!\!\!\sum_{k,l,m,n,p\atop\pm_1...\pm_5}\!\!\!\!\!\left(Q^{\pm_1...\pm_5}_{klmnp}\hat{\rho}_{00}\hat{c}_{\pm_1k}^{\pm_1}\hat{c}_{\pm_2 l}^{\pm_2}\hat{c}_{\pm_3m}^{\pm_3}\hat{c}_{\pm_4n}^{\pm_4}\hat{c}_{\pm_5p}^{\pm_5}+H.c.	\right)\nonumber\\
%&&+\sum_{k,l,m\atop\pm_1\pm_2\pm_3}\left(C^{\pm_1\pm_2\pm_3}_{klm}\hat{\rho}_{00}\hat{c}_{\pm_1 k}^{\pm_1}\hat{c}_{\pm_2 l}^{\pm_2}\hat{c}_{\pm_3 m}^{\pm_3}+H.c.	\right)\\
%\hat{\rho}_{10}&=&\sum_{k,l\atop \pm_1,\pm_2}\text{ }\left(A^{\pm_1\pm_2}_{kl}\hat{\rho}_{00}\hat{c}_{\pm_1k}^{\pm_1}\hat{c}_{\pm_2 l}^{\pm_2}+H.c.	\right)
%\end{eqnarray}
%The term $\hat{\rho}_{11}$ has a similar form to $\hat{\rho}_{01}$, but with different co-efficients. The evaluation of the coefficients $A^{\pm_1\pm_2}_{kl}$, $C^{\pm_1\pm_2\pm_3}_{klm}$  and $Q^{\pm_1...\pm_5}_{klmnp}$ is straightforward but lengthy, and will be reported in detail elsewhere \cite{tbppra}.

To interpret our perturbatively obtained stationary quantum state (\ref{eq:rho}), we compute the single particle reduced density operator $\hat{X}$, which acts in a single-particle Hilbert space:
\begin{eqnarray}
[\hat{X}]_{ml}=\frac{1}{N}\mathrm{Tr}\left(\hat{\rho}\hat{a}^{\dagger}_m\hat{a}_l	\right)\; \text{0}\leq\text{m,l}\leq\text{2M}\label{eq:rdm}\;.
\end{eqnarray}
Again we expand in $\gamma$ and $1/\sqrt{\bar{n}}$:
\begin{eqnarray}\label{}
	\hat{X}&=&\frac{1}{N}\sum_{ij}\hat{X}_{ij}(\gamma\bar{n})^{i}\,\bar{n}^{-j/2}\;.
\end{eqnarray}
Note that the Bogoliubov expansion of $\hat{a}_{m}$ means that different orders of $\hat{\rho}_{ij}$ contribute to the same order in $\hat{X}$. It is now straightforward to show that $\hat{X}_{00}=1/(2M)$ and $\hat{X}_{01}=0$. To zeroth order in $\gamma\bar{n}$, therefore, $\hat{X}$ has one constant eigenvector with eigenvalue of order 1 (the condensate), and many eigenvectors with eigenvalues of order $1/N$ (the non-condensate). 

 By introducing the compact notation
\begin{eqnarray}\label{}
	\frac{1}{\sqrt{2M}}\sum_{k=1}^{2M-1}e^{i\frac{\pi k m}{M}}\hat{b}_{k}=:\hat{\beta}_{m}
\end{eqnarray}
%(this is not a canonical transformation because $2M-1$ canonical $\hat{b}_{k}$ are spread into $2M$ $\hat{\beta}_{m}$), 
we can express the leading term involving the rate of heat exchange with the reservoirs as
%\begin{eqnarray}\label{}
%	\hat{a}_{m}=\frac{\hat{A}}{\sqrt{N}}  \left[\left(\bar{n}-\sum_{l=1}^{2M}\!\hat{\beta}^{\dagger}_l\hat{\beta}_l\right)^{1/{2}}+ \hat{\beta}_m\right]\;.
%\end{eqnarray}
\begin{eqnarray}\label{}
\hat{X}_{10} &=& \hat{X}_{10}^{NC}+\hat{X}_{10}^{C}\nonumber\\
\left[\hat{X}_{10}^{NC}\right]_{ml} &=& \mathrm{Tr}[\hat{\rho}_{10}(\hat{\beta}^{\dagger}_{m}\hat{\beta}_{l}-\sum_{k}\hat{\beta}_{k}^{\dagger}\hat{\beta}_{k})]\nonumber\\
\left[\hat{X}_{10} ^{C}\right]_{ml}&=& \mathrm{Tr}[\hat{\rho}_{11}(\hat{\beta}^{\dagger}_{m}+\hat{\beta}_{l})]\;.
\end{eqnarray}
There is no contribution from $\mathrm{Tr}\hat{\rho}_{12}$ because by normalizing $\mathrm{Tr}\hat{\rho}_{00}=1$ we have forced all higher $\hat{\rho}_{ij}$ to be traceless.)

Given the forms of the eigenvalues of $\hat{X}$ to zeroth order in $\gamma\bar{n}$, one can easily show that, to order $\gamma\bar{n}$, $\hat{X}_{10}^{NC}$ will perturb only the small eigenvalues of $\hat{X}$ and their eigenvectors, while $\hat{X}_{10}^{C}$ will alter only the condensate number and wave function by adding a homogeneous phase gradient representing a coherent condensate flow. Although the explicit evaluation of $\hat{X}_{10}$ is lengthy, therefore, it permits unambiguous analytical computation of the particle and energy currents, with the contributions to each from condensate and non-condensate fractions. 

The particle current can be defined by recognizing the discretized version of the continuity equation:
\begin{eqnarray}
\frac{d}{dt}\hat{n}_m&=& \frac{i}{\hbar}\left[\hat{H},\hat{n}_m	\right] = \hat{\jmath}_m - \hat{\jmath}_{m+1}, 
\label{eq:cont1}\\
\hat{\jmath}_m &:=& iJ\!\left(\hat{a}^{\dagger}_{m-1}\hat{a}_m - \hat{a}^{\dagger}_{m}\hat{a}_{m-1}\right)\;.
\label{eq:flowop}\end{eqnarray}
We observe that the expectation value $j_{m}=\mathrm{Tr}\hat{\rho}\hat{\jmath}_{m}$ is nothing but an off-diagonal element of the one-particle density operator $\hat{X}$, and so although we have only evaluated $\hat{X}$ perturbatively, no further approximation is made in extracting the particle current $j_{m}$ from it. 

Since we do not allow particle exchange with the reservoirs, there are no sinks or sources of particle density, and so our time-independent $\hat{\rho}$ must have uniform $j_{m} = j$, independent of $m$. And since our circular system's stationary state also has reflection symmetry through the diameter, this $j$ must vanish. By distinguishing the condensate and non-condensate fractions, however, we reveal internal convection. The non-condensate current can then be computed by
\begin{eqnarray}\label{}
	j_{m}^{NC}=2J\,\mathrm{Im}\left[\hat{X}^{NC}_{10}\right]_{(m-1),m}\;.
\end{eqnarray}
Apart from small distortions near $m=0,M$, $j_{m}^{NC}= I $ for $0<m<M$, and $j_{m}^{NC}= -I$ for $M<m<2M$. In other words, there is a uniform current $I$ of non-condensate particles from the hotter reservoir towards the cooler, around either side of the ring. From our explicit analytical results for $\hat{\rho}_{10}$ we obtain \cite{tbppra}
\begin{eqnarray}\label{Ieq}
I&=&\frac{J\gamma\bar{n}\Delta\beta}{(2M)^2}\sum_{k\neq l}^M\chi_{m-1,k},\chi_{m,l}(1-(-1)^{k+l})\nonumber\\
&&\qquad\qquad\cdot\frac{\omega_k+\omega_l}{\Omega_l^2-\Omega_k^2}(\Omega_kn_k+\Omega_ln_l)\;,\label{eq:flowresultsum}
\end{eqnarray}
where $\chi_{m,k}\!\propto\!\cos\frac{\pi km}{M}$ are Fourier factors, and $n_k$ is the expected quasiparticle population of mode $k$. 

While $\hat{X}_{10}^{NC}$ is obtained from $\hat{\rho}_{10}$, which involves only the Bogoliubov Hamiltonian $\hat{H}_{0}$, $\hat{X}_{10}^{C}$ is given by $\hat{\rho}_{11}$, which includes the post-Bogoliubov dynamics of $\hat{H}_{1}$. Because the total current $j$ vanishes exactly, however, we know that the condensate current balances the non-condensate:
\begin{eqnarray}\label{}
		j_{m}^{C}:=2J\,\mathrm{Im}\left[\hat{X}^{C}_{10}\right]_{(m-1),m} = - j_{m}^{NC}\;.
\end{eqnarray}
In this sense we have exhibited internal convection as a post-Bogoliubov effect, without needing to perform any post-Bogoliubov calculations (though these are nonetheless performed in \cite{tbppra}, and confirm the result). The only property of $\hat{H}_{1}$ that we have needed is its contribution to continuity. Furthermore, one can show from the fact that $\hat{H}$ is real that neither condensate nor non-condensate can have any stationary current of order $\gamma^{0}$ \cite{tbppra}. The $I$ given by (\ref{Ieq}) is thus the internal convection current.

%\begin{figure}
%\includegraphics[scale=0.8]{plot_final.jpg}
%\label{fig:plot}
%\caption{Continuum limit: keeping the length fixed the sum (\ref{eq:flowresultsum}) approaches the continuum limit result (\ref{eq:flowresult}) as the number of lattice sites M increases. Lattice effects: Using a small $k$ expansion of the system frequencies $\omega_k$, $\Omega_k$, the continuum limit  is only valid for nonvanishing interaction energy ($\frac{\kappa}{J}>1$). In such a regime, the total particle flow is proportional to $c^{-5}$.}
%\end{figure}

We can also derive $I$ for the continuum limit, by introducing a lattice spacing $a$, and keeping the system's length $2Ma$ fixed as $a\to0$. For a finite $I$, we must also extend our model straightforwardly by allowing each reservoir to couple to $L/a$ adjacent sites, for some fixed $L$ less than the total system size, rather than to only one site. For small relative temperature difference $\Delta T /T$, we obtain the continuum internal convection current \cite{tbppra}
\begin{eqnarray}
I_c&=&\frac{\pi^{5/2}}{120}\frac{\Delta T}{T} \left(\frac{k_{B}T}{\mu}\right)^{3}\frac{L}{\lambda_{f}}\rho\, c\label{eq:flowresult}\;,
\end{eqnarray}
where $\rho$ is the linear particle density, $c$ the speed of sound in the gas, $\mu\propto c^2$ the chemical potential, and $\lambda_{f}=(\sqrt{2}\rho\sigma)^{-1}$ the reservoir particles' mean free path in the sample gas (assuming all atoms have equal mass).  A numerical analysis shows that the discrete result (\ref{eq:flowresultsum}) is already well approximated by the continuum limit (\ref{eq:flowresult}) for system sizes around 100 lattice sites. 

Eqn.~(\ref{eq:flowresult}) is our main result. It includes no dependence on the total system size: ballistic transport in the collisionless regime means that the effect depends on the temperature difference, rather than the temperature gradient, as it would in the hydrodynamic regime. Currently, $k_{B}T \sim \mu$ is not difficult to attain experimentally, so it seems possible to achieve an internal convection condensate flow velocity $I/\rho$ not far below $c$, {\it i.e.} approaching the mm/s range, for sufficiently large condensates. The velocity of the much less dense normal component will be correspondingly greater, to obtain equal current. It should therefore be possible to observe internal convection as separation of condensate and thermal fraction in time of flight expansion.

L. G. is a recipient of a fellowship through funding of the Excellence Initiative (DFG/GSC 266).

\bibliography{bib.bib}
\end{document}